\documentclass[final]{IEEEtran}
\ifCLASSINFOpdf
\else
\fi

\usepackage{booktabs}
\usepackage{multirow}
\usepackage{graphicx}
\usepackage{xcolor}
\definecolor{darkblue}{rgb}{0.0,0.0,0.6}

\usepackage{wrapfig}

\begin{document}
%
\title{Sentiment Classification using N-gram IDF and Automated Machine Learning}

\author{
\IEEEauthorblockN{
Rungroj Maipradit\IEEEauthorrefmark{1},
Hideki Hata\IEEEauthorrefmark{1},
Kenichi Matsumoto\IEEEauthorrefmark{1}}\\
\IEEEauthorblockA{\IEEEauthorrefmark{1}Nara Institute of Science and Technology, Japan\\
\{maipradit.rungroj.mm6, hata, matumoto\}@is.naist.jp}
}


\maketitle

\begin{abstract}
We propose a sentiment classification method with a general machine learning framework.
For feature representation, n-gram IDF 
is used to extract software-engineering-related, dataset-specific, positive, neutral, and negative n-gram expressions. For classifiers, an automated machine learning tool 
is used.
In the comparison using publicly available datasets, our method achieved 
the highest F1 values in positive and negative sentences on all datasets.

{{\it keywords} Automated machine learning; N-gram IDF; Sentiment classification}
\end{abstract}


%
\IEEEpeerreviewmaketitle

\section{Introduction}

As software development is a human activity, identifying affective states in messages has become an important challenge to extract meaningful information. Sentiment analysis has been used to several practical purposes, such as
identifying problematic API features~\cite{6613842},
assessing the polarity of app reviews~\cite{Panichella:2015:IIM:2881297.2881394},
clarifying the impact of sentiment expressions to the issue resolution time~\cite{Ortu:2015:BMP:2820518.2820555},
and so on.

Because of the poor accuracy of existing sentiment analysis tools trained with general sentiment expressions~\cite{Jongeling:2017:NRU:3135854.3135907}, recent studies have tried to customize such tools with software engineering datasets~\cite{lin2018sentiment}. 
However, it is reported that no tool is ready to accurately classify sentences to negative, neutral, or positive, even if tools are specifically customized for certain software engineering tasks~\cite{lin2018sentiment}.

One of the difficulties in sentiment classification is the limitation in a bag-of-words model or polarity shifting because of function words and constructions in sentences~\cite{Li:2010:SCP:1873781.1873853}. Even if there are positive single words, the whole sentence can be negative because of negation, for example.
For this challenge, Lin et al. adopted \texttt{Stanford CoreNLP}, a recursive neural network based approach that can take into account the composition of words~\cite{D13-1170}, and prepared a software-engineering-specific sentiment dataset from a Stack Overflow dump~\cite{lin2018sentiment}. Despite their large amount of effort on fine-grained labeling to the tree structure of sentences, they reported negative results (low accuracy)~\cite{lin2018sentiment}.

In this paper we propose a machine-learning based approach using n-gram features and an automated machine learning tool for sentiment classification.
Although n-gram phrases are considered to be informative and useful compared to single words, using all n-gram phrases is not a good idea because of the large volume of data and many useless features~\cite{Bespalov:2011:SCB:2063576.2063635}. To address this problem, we utilize n-gram IDF, a theoretical extension of Inverse Document Frequency (IDF) proposed by Shirakawa et al.~\cite{Shirakawa:2015:NIG:2736277.2741628}.
IDF measures how much information the word provides; but it cannot handle multiple words. N-gram IDF is capable of handling n-gram phrases; therefore, we can extract useful n-gram phrases.

Automated machine learning is an emerging research area targeting the progressive automation of machine learning.
Two important problems are known in machine learning: no single machine learning technique give the best result on all datasets, and hyperparameter optimization is needed.
Automated machine learning addresses these problems by running multiple classifiers and tries different parameters to optimize the performance.
In this study, we use auto-sklearn, which contains 15 classification algorithms (random forest, kernel SVM, etc.), 14 feature pre-processing solutions (PCA, nystroem sampler, etc.), and 4 data pre-process solutions (one-hot encoding, rescaling, etc.)~\cite{NIPS2015_5872}.
Using n-gram IDF and auto-sklearn tools, Wattanakriengkrai et al. outperformed the state-of-the-art self-admitted technical debt identification~\cite{id1275}.

\begin{figure*}
    \centering
    \includegraphics[width=.8\textwidth]{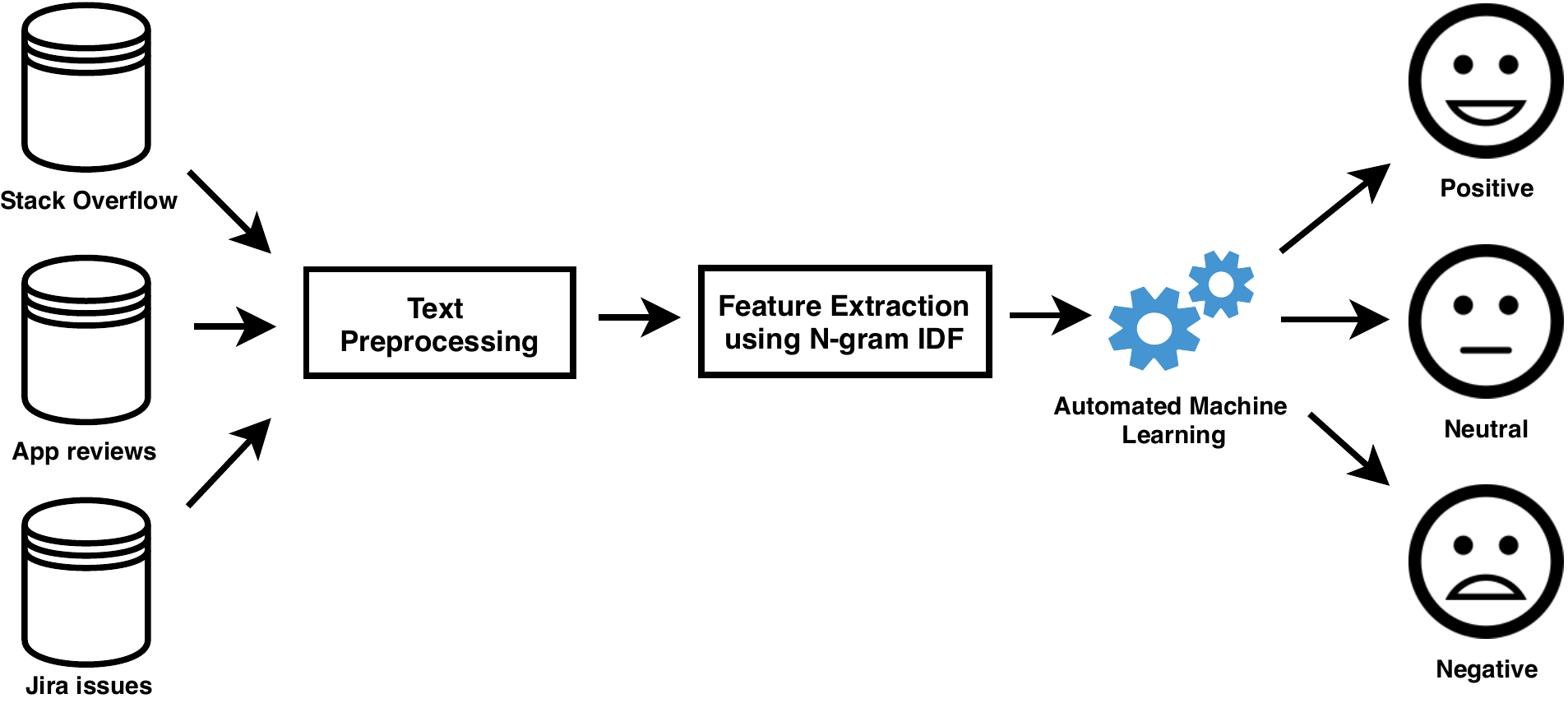}
    \caption{An overview of our sentiment classification approach}
    \label{fig:overview}
\end{figure*}

\section{Method}

Figure~\ref{fig:overview} shows an overview of our method with the following three components. 

\textbf{Text Preprocessing}.
Messages in software document sometimes contain special characters.
We remove characters that are neither English characters nor numbers.
Stop words are also removed by using spaCy library. 
spaCy tokenizes text and finds part of speech and tag of each token and also checks whether the token appears in the stop word list.

\textbf{Feature extraction using N-gram IDF}.
N-gram IDF is a theoretical extension of IDF for handling words and phrases of any length by bridging the gap between term weighting and multiword expression extraction. N-gram IDF can identify dominant n-grams among overlapping ones~\cite{Shirakawa:2015:NIG:2736277.2741628}.
In this study,
we use N-gram Weighting Scheme tool~\cite{Shirakawa:2015:NIG:2736277.2741628}.
The result after applying this tool is a dictionary of n-gram phrases ($n\leq10$ as a default setting) with their frequencies.
N-gram phrases appear only one time in the whole document (frequency equal one) are removed, since they are not useful for training.

\textbf{Automated machine learning}.
To classify sentences into positive, neutral, and negative, we use auto-sklearn,
an automated machine learning tool.
Automated machine learning tries running multiple classifiers and applying different parameters to derive better performances.
Auto-sklearn composes two steps: meta-learning and automated ensemble construction~\cite{NIPS2015_5872}.
We ran auto-sklearn with 64 gigabytes of memories, set 90 minutes limitation for each round, and configure it to optimize for a weighted F1 value, an average F1 value for three classes weighted by the number of true instance of each class.

\section{Evaluation}
\subsection{Datasets and Settings}
We employ a dataset provided by Lin et al.'s  work~\cite{lin2018sentiment}.
There are three types of document in the dataset;
sentences in questions and answers on Stack Overflow,
reviews of mobile applications,
and comments on Jira issue trackers.
Each dataset has texts and labels of positive, neutral and negative.

Since our method requires just labeled (positive, neutral, or negative) sentences, we can use dataset-specific data for training.
For training and testing, we apply 10-fold cross-validation for each dataset, that is,
we split sentences in a dataset into ten subsets with maintaining the ratio from the oracles by using the function StratifiedShuffleSplit in scikit-learn. 

\subsection{Sentiment Classification Tools}
To assess the performance of our method,
we compare our method with tools presented in the previous work~\cite{lin2018sentiment}. \\
    \textbf{SentiStrength} estimates the strength of positive and negative scores based on the sentiment word strength list prepared from MySpace comments~\cite{Thelwall:2010:SSD:1890706.1890713}. \\ 
    \textbf{NLTK} is a natural language toolkit and is able to do sentiment analysis based on lexicon and rule-based VADER (Valence Aware Dictionary and sEntiment Reasoner), which is specifically tuned for sentiments expressed in social media~\cite{ICWSM148109}. \\
    \textbf{Stanford CoreNLP} adopts a deep learning model to compute the sentiment based on how words compose the meaning of the sentence~\cite{D13-1170}. 
    The model has been trained on movie reviews. \\
    \textbf{SentiStrength-SE} is a tool build on top of SentiStrength and has been trained on JIRA issue comments~\cite{Islam:2017:LAS:3104188.3104215}. \\ 
    \textbf{Stanford CoreNLP SO} prepared Stack Overflow discussions to train a model of Standford CoreNLP~\cite{lin2018sentiment}.


\begin{table*}
\centering
\caption{The comparison result of the number of corrected prediction, precision, recall, and f1-score} 
\label{tab:result}
\begin{tabular}{llr|rrr|rrr|rrr}
\toprule
\multirow{2}{*}{\textbf{dataset}} & \multirow{2}{*}{\textbf{tool}} & \multicolumn{1}{c|}{\multirow{2}{*}{\textbf{\begin{tabular}[c]{@{}c@{}}\# correct\\ prediction\end{tabular}}}} & \multicolumn{3}{c|}{\textbf{positive}}& \multicolumn{3}{c|}{\textbf{neutral}} & \multicolumn{3}{c}{\textbf{negative}} \\ 
\multicolumn{1}{c}{} & \multicolumn{1}{c}{} & \multicolumn{1}{c|}{} & \multicolumn{1}{c}{\textbf{precision}} & \multicolumn{1}{c}{\textbf{recall}} & \multicolumn{1}{c|}{\textbf{F1}} & \multicolumn{1}{c}{\textbf{precision}} & \multicolumn{1}{c}{\textbf{recall}} & \multicolumn{1}{c|}{\textbf{F1}} & \multicolumn{1}{c}{\textbf{precision}} & \multicolumn{1}{c}{\textbf{recall}} & \multicolumn{1}{c}{\textbf{F1}} \\
\midrule
\textbf{Stack Overflow} & SentiStrength & 1043 & 0.200 & \textbf{0.359} & 0.257 & 0.858 & 0.772 & 0.813 & 0.397 & 0.433 & 0.414 \\
 positive: 178 & NLTK & 1168 & 0.317 & 0.244 & 0.276 & 0.815 & \textbf{0.941} & 0.873 & \textbf{0.625} & 0.084 & 0.148 \\
 neutral: 1,191 & Standford CoreNLP & 604 & 0.231 & 0.344 & 0.276 & \textbf{0.884} & 0.344 & 0.495 & 0.177 & \textbf{0.837} & 0.292 \\
 negative: 131 & SentiStrength-SE & 1170 & 0.312 & 0.221 & 0.259 & 0.826 & 0.930 & 0.875 & 0.500 & 0.185 & 0.270 \\
 sum: 1,500 & Stanford CoreNLP SO & 1139 & 0.317 & 0.145 & 0.199 & 0.836 & 0.886 & 0.860 & 0.365 & 0.365 & 0.365 \\
 & N-gram auto-sklearn & \textbf{1317} & \textbf{0.667} & 0.316 & \textbf{0.418} & 0.871 & 0.939 & \textbf{0.904} & 0.600 & 0.472 & \textbf{0.514} \\
 \cmidrule{2-12}
 & N-gram auto-sklearn with SMOTE$\dagger$ & - & 0.680 & 0.005 & 0.009 & 0.344 & 0.930 & 0.499 & 0.657 & 0.160 & 0.251 \\
\midrule
\textbf{App reviews} & SentiStrength & 213 & 0.745 & 0.866 & 0.801 & 0.113 & 0.320 & 0.167 & 0.815 & 0.338 & 0.478 \\
 positive: 186 & NLTK & 184 & 0.751 & 0.812 & 0.780 & 0.093 & \textbf{0.440} & 0.154 & \textbf{1.000} & 0.169 & 0.289 \\
 neutral: 25 & Standford CoreNLP & 237 & 0.831 & 0.715 & 0.769 & \textbf{0.176} & 0.240 & \textbf{0.203} & 0.667 & 0.754 & 0.708 \\
 negative: 130 & SentiStrength-SE & 201 & 0.741 & 0.817 & 0.777 & 0.106 & 0.400 & 0.168 & 0.929 & 0.300 & 0.454 \\
 sum: 341 & Stanford CoreNLP SO & 142 & 0.770 & 0.253 & 0.381 & 0.084 & 0.320 & 0.133 & 0.470 & 0.669 & 0.552 \\
 & N-gram auto-sklearn & \textbf{293} & \textbf{0.822} & \textbf{0.894} & \textbf{0.853} & 0.083 & 0.066 & 0.073 & 0.823 & \textbf{0.808} & \textbf{0.807} \\
 \cmidrule{2-12}
 & N-gram auto-sklearn with SMOTE$\dagger$ & - & 0.520 & 0.885 & 0.641 & 0.100 & 0.058 & 0.073 & 0.648 & 0.622 & 0.607 \\
\midrule
\textbf{Jira issues} & SentiStrength & 714 & 0.850 & \textbf{0.921} & 0.884 & - & - & - & 0.993 & 0.703 & 0.823 \\
 positive: 290 & NLTK & 276 & 0.840 & 0.362 & 0.506 & - & - & - & \textbf{1.000} & 0.269 & 0.424 \\
 neutral: 0 & Standford CoreNLP & 626 & 0.726 & 0.621 & 0.669 & - & - & - & 0.945 & 0.701 & 0.805 \\
 negative: 636 & SentiStrength-SE & 704 & 0.948 & 0.883 & 0.914 & - & - & - & 0.996 & 0.704 & 0.825 \\
 sum: 926 & Stanford CoreNLP SO & 333 & 0.635 & 0.252 & 0.361 & - & - & - & 0.724 & 0.409 & 0.523 \\
 & N-gram auto-sklearn & \textbf{884} & \textbf{0.960} & 0.839 & \textbf{0.893} & - & - & - & 0.932 & \textbf{0.982} & \textbf{0.956} \\
 \cmidrule{2-12}
 & N-gram auto-sklearn with SMOTE$\dagger$ & - & 0.986 & 0.704 & 0.809 & - & - & - & 0.781 & 0.988 & 0.872 \\
\bottomrule
\multicolumn{12}{r}{$\dagger$ Applying SMOTE, a oversampling technique, for our method.} \\
\end{tabular}
\end{table*}

\subsection{Result}
Table~\ref{tab:result} shows
the number of correct predictions, precision, recall, and F1 values with all tools including our method (\texttt{n-gram auto-sklearn}). These values were presented in~\cite{lin2018sentiment} (F1 values are calculated by us).
Precision, recall, and F1 values are derived as the average from the 10 rounds of our 10-fold cross-validation, since same data can appear in different rounds with StratifiedShuffleSplit.

We can see that the number of correct predictions are higher with our method in all three datasets, and our method achieved the highest F1 values for all three positive, all three negative, and one neutral.
Although the values are low for the neutral class in App reviews, this is because
the amount of neutral sentences is small in this dataset.
In summary, our method using n-gram IDF and automated machine learning (auto-sklearn) largely outperformed existing sentiment analysis tools.
Since our method relies on n-gram phrases, it cannot properly classify text without known n-gram phrases. Although a negative sentence ``They all fail with the following error'' was correctly classified with SentiStrength, NLTK, and Stanford CoreNLP, our method classified as neutral. Preparing more data is preferable to improve the performance.

Note that only our method trains within-dataset for all three cases. Although within-dataset training can improve the performances of other tools,  preparing training data for those sentiment analysis tools require considerable manual effort~\cite{lin2018sentiment}. Since our method can automatically learn dataset-specific text features, learning within-dataset is practically feasible.

The following are classifiers achieved the top three performances in \texttt{auto-sklearn} for each dataset.
\begin{itemize}
    \item \textbf{Stack Overflow}: Linear Discriminant Analysis, LibSVM Support Vector Classification, and Liblinear Support Vector Classification
    \item \textbf{App reviews}: Random forest, LibSVM Support Vector Classification, and Naive Bayes classifier for multinomial models
    \item \textbf{Jira issues}: Naive Bayes classifier for multinomial models, Adaptive Boosting, and Linear Discriminant Analysis
\end{itemize}
If we have a new unlabeled dataset, we can first try one of the common classifiers. 
By manually annotating labels, we can try auto-sklearn to find the best classifier for the dataset.



\section{Discussions}

\subsection{Threats to Validity}
\textbf{Competitive sentiment classification tools are trained only with specific dataset}.
Since our method is based on a general text classification approach, we could conduct within-dataset training.
However, because of a considerable amount of manual effort for training sentiment classification tools, they had been trained only with specific datasets.
Although we think this is an advantage of our method, the comparison is not with the same condition.

\textbf{Imbalanced data}.
In this multi-class classification, some datasets are not balanced; neutral class is the majority for Stack Overflow and no neural data for Jira issues. Applying some balancing techniques may improve the overall performances.

\textbf{Our study might not be generalize to other datasets}.
Our approach is applied to comments, reviews, and questions and answers. Other types of document related to software engineering may derive different results.


\begin{table*}
\centering
\caption{Obtained n-gram phrases (selected)}
\label{tab:n-gram}
\begin{tabular}{l|l|l|l}
\toprule
\textbf{dataset} & \textbf{positive} & \textbf{neutral} & \textbf{negative} \\
\midrule
\multirow{5}{*}{\textbf{Stack Overflow}} & `useful' & `suggest', `using' & `wrong' \\
 & `the', `simplest', `solution' & `everyone' & `bug' \\
 & `more', `efficient', `to' & `limitation' & `i', `do', `n', `t', `understand' \\
 & `helpful' & `appointment' & `i', `do', `n', `t', `know' \\
 & `a', `good', `example' & `technical' & `does', `n', `t', `work' \\
\midrule
\multirow{5}{*}{\textbf{App reviews}} & `thanks', `for' & `let', `me', `know' & `impossible', `to', `use' \\
 & `really', `like', `this', `app' & `gets', `the', `job', `done' & `game', `constantly', `freezes' \\
 & `well', `done' & `android', `device' & `uninstalled' \\
 & `awesome' & `allow' & `disappointing' \\
 & `easy', `to', `use', `and' & `suggestions' & `lack', `of', `features' \\
\midrule
\multirow{5}{*}{\textbf{Jira issues}} & `thank', `you', `very', `much' & - & `problems' \\
 & `looks', `good', `we' & - & `is', `bad' \\
 & `awesome', `work' & - & `i', `disagree' \\
 & `thanks', `for', `your', `help' & - & `really', `sucks' \\
 & `awesome', `stuff' & - & `this', `bug' \\
\bottomrule
\end{tabular}
\end{table*}

\subsection{Obtained N-gram Phrases}
Why our method achieved high accuracy performance in sentiment classification?
Table~\ref{tab:n-gram} shows selected n-gram phrases, which were useful for classifying positive, neutral, and negative statements, obtained in each dataset.
For negative, we see `bug', a software-engineering-specific negative word, and many negation expressions. We can also see reasonable n-gram phrases for positive cases, such as `useful', `really like this app', `awesome work', and so on.
We can think that because of these dataset-specific positive, neutral, and negative patterns, n-gram IDF worked well for resolving the limitation in a bag-of-words model and our method have resulted in good performance.


\section{Conclusion}
In this paper, we proposed a sentiment classification method using n-gram IDF and automated machine learning.
We apply this method on three datasets including question and answer from Stack Overflow, reviews of mobile applications, and comments on Jira issue trackers.


Our good classification performance is not based only on an advanced automated machine learning. N-gram IDF also worked well to capture dataset-specific, software-engineering-related positive, neutral, and negative expressions.
Because of the capability of extracting useful sentiment expressions with n-gram IDF, our method can be applicable to various software engineering datasets.





%

\begin{thebibliography}{10}
\providecommand{\url}[1]{#1}
\csname url@samestyle\endcsname
\providecommand{\newblock}{\relax}
\providecommand{\bibinfo}[2]{#2}
\providecommand{\BIBentrySTDinterwordspacing}{\spaceskip=0pt\relax}
\providecommand{\BIBentryALTinterwordstretchfactor}{4}
\providecommand{\BIBentryALTinterwordspacing}{\spaceskip=\fontdimen2\font plus
\BIBentryALTinterwordstretchfactor\fontdimen3\font minus
  \fontdimen4\font\relax}
\providecommand{\BIBforeignlanguage}[2]{{%
\expandafter\ifx\csname l@#1\endcsname\relax
\typeout{** WARNING: IEEEtran.bst: No hyphenation pattern has been}%
\typeout{** loaded for the language `#1'. Using the pattern for}%
\typeout{** the default language instead.}%
\else
\language=\csname l@#1\endcsname
\fi
#2}}
\providecommand{\BIBdecl}{\relax}
\BIBdecl

\bibitem{6613842}
Y.~Zhang and D.~Hou, ``Extracting problematic API features from forum
  discussions,'' in \emph{Proceedings of 21st International Conference on Program
  Comprehension (ICPC)}, 2013, pp. 142--151.

\bibitem{Panichella:2015:IIM:2881297.2881394}
S.~Panichella, A.~Di~Sorbo, E.~Guzman, C.~A. Visaggio, G.~Canfora, and H.~C.
  Gall, ``How can i improve my app? classifying user reviews for software
  maintenance and evolution,'' in \emph{Proceedings of 31st IEEE
  International Conference on Software Maintenance and Evolution (ICSME)},
  2015, pp. 281--290.

\bibitem{Ortu:2015:BMP:2820518.2820555}
M.~Ortu, B.~Adams, G.~Destefanis, P.~Tourani, M.~Marchesi, and R.~Tonelli,
  ``Are bullies more productive?: Empirical study of affectiveness vs. issue
  fixing time,'' in \emph{Proceedings of 12th Working Conference on Mining
  Software Repositories (MSR)}, 2015, pp. 303--313.

\bibitem{Jongeling:2017:NRU:3135854.3135907}
R.~Jongeling, P.~Sarkar, S.~Datta, and A.~Serebrenik, ``On negative results
  when using sentiment analysis tools for software engineering research,''
  \emph{Empirical Software Engineering}, vol.~22, no.~5, pp. 2543--2584, Oct. 2017.

\bibitem{lin2018sentiment}
B.~Lin, F.~Zampetti, G.~Bavota, M.~Di~Penta, M.~Lanza, and R.~Oliveto,
  ``Sentiment analysis for software engineering: How far can we go?'' in
  \emph{Proceedings of 40th International Conference on Software
  Engineering (ICSE)}, 2018, pp. 94--104.

\bibitem{Li:2010:SCP:1873781.1873853}
S.~Li, S.~Y.~M. Lee, Y.~Chen, C.-R. Huang, and G.~Zhou, ``Sentiment
  classification and polarity shifting,'' in \emph{Proceedings of 23rd
  International Conference on Computational Linguistics (COLING)},
  2010, pp. 635--643.

\bibitem{D13-1170}
R.~Socher, A.~Perelygin, J.~Wu, J.~Chuang, C.~D. Manning, A.~Ng, and C.~Potts,
  ``Recursive deep models for semantic compositionality over a sentiment
  treebank,'' in \emph{Proceedings of 2013 Conference on Empirical Methods
  in Natural Language Processing (EMNLP)}, 2013, pp. 1631--1642.

\bibitem{Bespalov:2011:SCB:2063576.2063635}
D.~Bespalov, B.~Bai, Y.~Qi, and A.~Shokoufandeh, ``Sentiment classification
  based on supervised latent n-gram analysis,'' in \emph{Proceedings of
  20th ACM International Conference on Information and Knowledge Management (CIKM)},
  2011, pp. 375--382.

\bibitem{Shirakawa:2015:NIG:2736277.2741628}
M.~Shirakawa, T.~Hara, and S.~Nishio, ``N-gram IDF: A global term weighting
  scheme based on information distance,'' in \emph{Proceedings of 24th
  International Conference on World Wide Web (WWW)}, 2015, pp.
  960--970.

\bibitem{NIPS2015_5872}
M.~Feurer, A.~Klein, K.~Eggensperger, J.~Springenberg, M.~Blum, and F.~Hutter,
  ``Efficient and robust automated machine learning,'' in \emph{Proceedings of 
  28th International Conference on Neural Information Processing Systems (NIPS)}, 2015, pp. 2962--2970.

\bibitem{id1275}
S.~Wattanakriengkrai, R.~Maipradit, H.~Hata, M.~Choetkiertikul, T.~Sunetnanta,
  and K.~Matsumoto, ``Identifying design and requirement self-admitted
  technical debt using n-gram IDF,'' in \emph{Proceedings of 9th IEEE International
  Workshop on Empirical Software Engineering in Practice (IWESEP)}, 2018,
  pp. 7--12.

\bibitem{Thelwall:2010:SSD:1890706.1890713}
M.~Thelwall, K.~Buckley, G.~Paltoglou, D.~Cai, and A.~Kappas, ``Sentiment
  strength detection in short informal text,'' \emph{Journal of the American Society for Information Science and Technology}, vol.~61, no.~12, pp. 2544--2558, Dec. 2010.

\bibitem{ICWSM148109}
C.~Hutto and E.~Gilbert, ``Vader: A parsimonious rule-based model for sentiment
  analysis of social media text,'' in \emph{Proceedings of 8th International AAAI Conference on Weblogs and Social Media}, 2014, pp. 216--225.

\bibitem{Islam:2017:LAS:3104188.3104215}
M.~R. Islam and M.~F. Zibran, ``Leveraging automated sentiment analysis in
  software engineering,'' in \emph{Proceedings of 14th International
  Conference on Mining Software Repositories (MSR)}, 2017, pp.
  203--214.

\end{thebibliography}

\end{document}